# Discovery of new drug therapeutic indications from gene mutation information for hepatocellular carcinoma

Liang Yu[1]*, Fengdan Xu[1], Lin Gao[1]

[1]School of Computer Science and Technology, Xidian University, Xi'an, PR China

*Corresponding author

E-mail: lyu@xidian.edu.cn(LY)

## Abstract

Hepatocellular carcinoma (HCC) is the most common primary liver malignancy and is a leading cause of cancer-related death worldwide. However, cure is not possible with currently used therapies, and there is not so much approved targeted therapy for HCC despite numerous attempts and clinical trials. So, it is essential to identify additional therapeutic strategies to block the growth of HCC tumors. As a cancer disease, it is associated with aberrant genomic and transcriptional landscapes. We sought to use a systematic drug repositioning bioinformatics approach to identify novel candidate drugs to treat HCC, which considers not only aberrant genomic information, but also the changes of transcriptional landscapes. First, we screen the collection of HCC feature genes that frequently mutated in most samples of HCC based on human mutation data. Then, the gene expression data of HCC in TCGA are combined to classify the kernel genes of HCC. Finally, the therapeutic score (*TS*) of each drug is calculated based on the kolmogorov-smirnov statistical method. Using this strategy, we identified five drugs that associated with HCC, including three drugs that could treat HCC and two drugs that might have side-effect on HCC. In addition, we also make Connectivity Map (CMap) [18] profiles similarity analysis and KEGG enrichment analysis on drug targets. All these findings suggest that our approach is effective for accurate discovering novel therapeutic options for HCC and easily to be extended to other tumors.

## Introduction

The identification of therapeutic approaches for the treatment of cancer is an arduous, costly, and often inefficient process[1]. Drug repositioning, which is the discovery of new indications for existing drugs that are outside their original indications, is an increasingly attractive mode of new use discovery. In addition to saving time and money, an advantage of drug repurposing approaches is the fact that existing drugs have already been vetted in terms of safety, dosage, and toxicity[2]. Therefore, repurposed drugs can often enter clinical trials much more rapidly than newly developed drugs. Rapid advances in genomics have led to the generation of large volumes of genomic and transcriptomic data for a diverse set of disease samples, normal tissue samples, animal models and cell lines. Transcriptomic profiles, such as gene expression data are most widely used for drug repositioning. One key source of data behind several repurposing efforts is the Connectivity Map (CMap)[3], which produced large-scale gene expression profiles from human cancer cell lines treated with different drug compounds under different conditions. The CMap approach attempts to take a more holistic view of these transcriptomic data and use them to connect expression profiles across conditions[3]. In particular, it indicates that drugs could potentially have a therapeutic effect on the disease if there is a strong negative correlation between the disease signatures and the drug expression profiles. For example, by systematically comparing gene expression signatures of inflammatory bowel disease (IBD) derived from GEO against a set of drug gene expression signatures comprising 164 drug compounds from CMap, Dudley et al.[4] inferred several new interesting drug–disease pairs and validated one pair in IBD preclinical models. In another case, Jahchan et al.[5] used a similar systematic drug-repositioning bioinformatics approach to query a large compendium of gene expression profiles to identify antidepressant drugs for the treatment of small cell lung cancer. A growing body of literature supports the use of CMap for drug repositioning; however, there are problems as well. A candidate can often be strengthened using independent disease signatures. But disease signatures are often selected by statistical methods, they are lack of biological information.

Hepatocellular carcinoma (HCC) is the most common primary liver malignancy and is a leading cause of cancer-related death worldwide[6]. HCC is often induced by infections of hepatitis virus B (HBV) [7] and C (HCV) [8], exposure to aflatoxin B1 from Aspergillus [9], alcohol abuse [10], or from non-alcoholic steatohepatitis [11]. HCC is now the third leading cause of cancer deaths worldwide. However, cure is not possible with currently used therapies, and there is not so much approved targeted therapy for HCC despite numerous attempts and clinical trials. So, it is essential to identify additional therapeutic strategies to block the growth of HCC tumors.

Many diseases, but very especially cancer, are associated with aberrant genomic and transcriptional landscapes [12]. In this study, we sought to use a systematic drug repositioning bioinformatics approach to identify novel candidate drugs to treat HCC. First, we screen the collection of HCC feature genes that frequently mutated in most samples of HCC based on human mutation data. Then, the gene expression data of HCC in TCGA are combined to classify the gene set of HCC. Finally, the therapeutic score (*TS*) of each drug is calculated based on the kolmogorov-smirnov statistical method. Using this strategy, we identified five drugs that associated with HCC, including three drugs that could cure HCC and two drugs that might have bad effect on HCC. In addition, we also make CMap [3] profiles similarity analysis and KEGG enrichment analysis on drug targets. All these findings suggest that our approach is effective for accurate discovering novel therapeutic options for HCC and easily to be extended to other tumors.

## Results

**Analysis of disease characteristics of HCC**

We characterized kernel, secondary, and marginal genes in the context of protein interaction (PPIs) network, PubMed (www.ncbi.nlm.nih.gov/pubmed) [13], and Gene Ontology [14] term annotation. The Human Protein Reference Database (HPRD) [15] is a resource for experimentally derived information about the human proteome including protein–protein interactions, post-translational modifications (PTMs) and other information. We downloaded all human PPIs from this database, containing 15231 proteins and 38,167 interactions. Interestingly, we found that all of the three gene types have a heterogeneous degree distribution, and kernel genes tend to have higher degrees compared with secondary and marginal genes (Fig.1 A). Similarly, kernel genes are associated with more PubMed recordsand Gene Ontology term annotation than secondary and marginal genes (Fig.1 B and C).

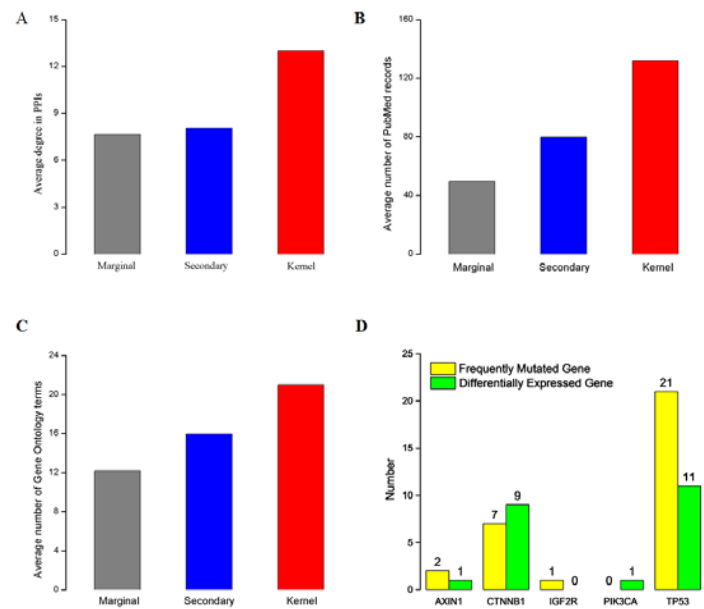

Figure 1. Characteristics of the three gene types. A. Average degree for three different gene types. B. Average PubMed records associated for each gene type. C. Gene Ontology terms annotated for each gene type. In figure A, B, and C, the red rectangle represents kernel genes, the blue rectangle represents secondary genes, the gray rectangle represents marginal genes. In figure D, green rectangle represents differentially expressed genes in HCC, and yellow rectangle represents frequently mutated genes in patients with HCC.

**Table 1. HCC related genes extracted from OMIM.**

| Gene Names | Gene Entrez IDs |
|:---:|:---:|
| IGF2R | 3482 |
| CASP8 | 841 |
| MET | 4233 |
| PDGFRL | 5157 |
| TP53 | 7157 |
| PIK3CA | 5290 |
| LCO | 3935 |
| CTNNB1 | 1499 |
| AXIN1 | 8312 |

In order to analyze biological functions of kernel genes, we analysis the nine HCC pathogenic genes obtained from Online Mendelian Inheritance in Man (OMIM) [16] from two aspects of gene mutation and expression level change. These nine HCC pathogenic genes (Table 1) are IGF2R, CASP8, MET, PDGFRL, TP53, PIK3CA, LCO, CTNNB1, and AXIN1. We find that five (IGF2R, TP53, PIK3CA, CTNNB1, AXIN1) of these genes are belong to kernel genes, these genes are frequent mutations, but their expression level don't change significantly. For direct neighbors in PPIs of these five genes, we find that there are frequently mutated or differentially expressed genes (see Fig.1 D) among their direct neighbors. TP53 is a quite important tumor suppressor gene, it can translate and synthesize P53 protein, and P53 protein is a vital regulator for cell growth, proliferation and injury repair. For the direct neighbors of TP53, there are 27 frequently mutated genes, and 11 differentially expressed genes. CTNNB1 gene can encode β-catenin, a dual function protein that involves in regulation and coordination of cell–cell adhesion and gene transcription [17]. Recent study of HCC has shown that CTNNB1 gene mutations and overexpression of its encoded protein are closely related to occurrence, progression and prognosis of tumor [18]. CTNNB1 has 7 frequently mutated direct neighbors, and 9 differentially expressed direct neighbors. The above statistics show that the kernel genes that are screened by mutation and gene expression data contain more comprehensive biological information and can characterize the disease characteristics of HCC.

**Choose Potential HCC Drugs through CTD Benchmark**

To find most likely HCC-related drugs, we need evaluate the precision of our method firstly. We take Comparative Toxicogenomics Database (CTD) [19] as benchmark. CTD provides manually curated information about drug-gene interactions, drug-disease and gene-disease relationships. Curated chemical-disease associations are extracted from the published literature by CTD biocurators and inferred associations are established via CTD curated chemical-gene interactions.

For a drug in CMap, if it cannot find corresponding chemical name in CTD, we will not calculate its therapeutic score (defined in section "Methods"). In this way, we finally get 1168 scored drugs. Because most drug-disease associations in CTD are not marked as positive or negative, we rank the 1168 drugs in descending order by the absolute values of their therapeutic scores. We know the top drugs imply stronger connections with HCC. And then we calculate the precisions of our approach at different top-x drugs, which are shown in Fig 2. The precision is calculated as follows:

$$precision = \frac{P_{CTD}}{P} \qquad (1)$$

where $P$ represents the number of top-x drugs; $P_{CTD}$ represents the number of drugs in the top-x drugs can be found related with HCC in CTD database.

We find in the top-10 drugs (x=10), there are 9 drugs associated with HCC in CTD. That is to say, the precision is 0.9. For the top-20 drugs (x=20), the precision is 0.85 and there are three potentially HCC-related drugs. When x is 30, its precision is 0.83 and we get five potential drugs with HCC. From the Fig 2, we notice that with the increase of x, the precision declines and the number of potential drugs increases. We hope we can predict relatively more HCC-related drugs with high precision.

Theinferredore, we choose top-30 (x = 30) drugs for further analysis.

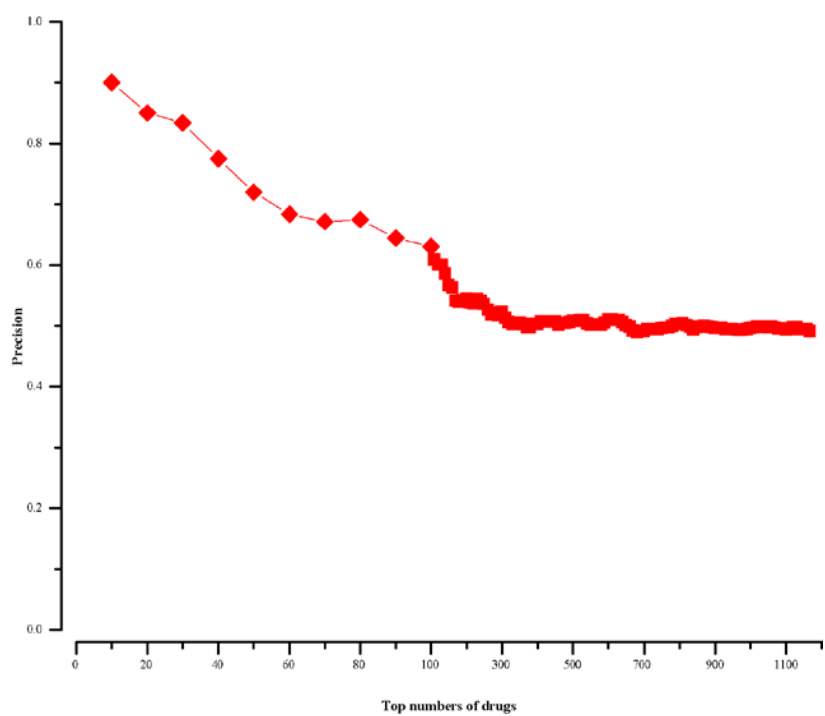

Fig 2. The precision of our approach at different top-x drugs.

**Validate potentially HCC-related Drugs through PubMed Literature**

In the above section, we choose the top-30 drugs (precision = 0.83) for further analysis. There are 19 therapeutic drugs with negative $TS$ values in the top-30 drugs, shown in Table 1. 16 of them can be found having connections with HCC in CTD [19]. Three of the 16 drugs are marked as therapeutic drug (Rank=1, Rank=9, Rank= 11 and Evidence = "T" in Table 2) for HCC .Meanwhile, one drug is marked as marker/mechanism drug (Rank=15, Evidence = "M" in Table 2) for HCC and the other 12 inferred drugs are unmarked in CTD. Here, we can indicate these 12 unmarked drugs are possibly therapeutic drugs for HCC. The rest three drugs (Securinine, Mercaptopurine and Reserpine) are newly predicted ones by our method, which are marked as bold in Table 1. Based on PubMed, we analyze the three drugs further. PubMed [13] is a free resource developed and maintained by the National Center for Biotechnology Information (NCBI) at the National Library of Medicine (NLM).

PubMed comprises more than 26 million inferrederences and abstracts on life sciences and biomedical topics.

**Table 2. Nineteen Therapeutic Drugs for HCC in the Top-30 Drugs.**

| Rank | Drug Name | Evidence | Inferred Count |
|------|-----------|----------|----------------|
| 1 | Daunorubicin | T | 42 |
| 2 | Chrysin | Inferred | 34 |
| 3 | Topiramate | Inferred | 8 |
| **4** | **Securinine** | **NULL** | **NULL** |
| 5 | Piperlongumine | Inferred | 8 |
| 6 | Luteolin | Inferred | 28 |
| 7 | Apigenin | Inferred | 36 |
| 8 | Celastrol | Inferred | 19 |
| 9 | Sirolimus | T | 68 |
| **10** | **Mercaptopurine** | **NULL** | **NULL** |
| 11 | Genistein | T | 93 |
| 12 | Irinotecan | Inferred | 46 |
| 13 | Sanguinarine | Inferred | 5 |
| 14 | Tyrphostin Ag-825 | Inferred | 7 |
| 15 | Decitabine | M | 84 |
| 16 | Camptothecin | Inferred | 28 |
| **17** | **Reserpine** | **NULL** | **NULL** |
| 18 | Mycophenolic Acid | Inferred | 7 |
| 19 | Tyrphostin Ag-1478 | Inferred | 35 |

Evidence represents a drug-disease association is curated, inferred or not existed in CTD database. Curated associations include three types: marker/mechanism (Evidence = "M"), therapeutic (Evidence = "T"), marker/mechanism & therapeutic (Evidence = "M&T"). If an association is inferred by CTD, Evidence = "inferred", and if it is not existed in CTD, Evidence = "NULL"; Inferred Count represents the number of inferrederence (s) for the curated and inferred associations. If an association is not existed in CTD, Inferred Count = "NULL".

Securinine (Rank=4 in Table 2), a quinolizine pseudoalkaloid (not from amino acid) from securinega suffurutiosa or securinini nitras, is one of central nervous stimulants and clinically applied to treat amyotrophic lateral sclerosis (ALS) [20], poliomyelitis [21] and multiple sclerosis [22]. It has been found to be active as a γ-amino butyric acid (GABA) receptor antagonist [23]. GABA is the chief inhibitory neurotransmitter in the central nervous system and plays a principal role in reducing neuronal excitability throughout the nervous system. Studies show that GABA stimulates HCC cell line HepG2 growth [24]. Consequently, it means that securinine is a promising agent with therapeutic potential for HCC through inhibiting GABA receptor. The mechanism is depicted in Fig 3A.

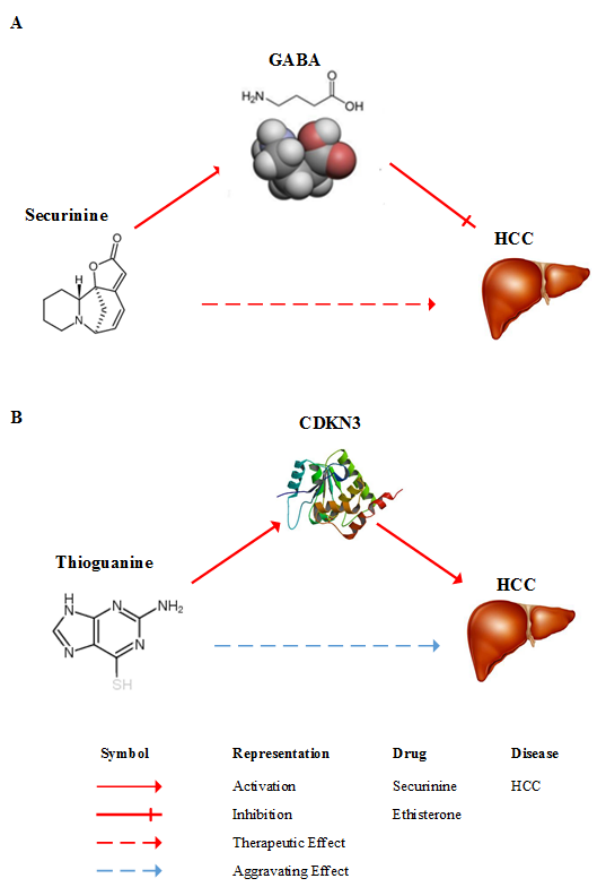

Fig 3. Description of the mechanism of drugs and HCC. A. The mechanism of securinine treating HCC. Securinine has been found to be active as a γ-amino butyric acid (GABA) receptor antagonist. GABA stimulates HCC cell line HepG2 growth. Consequently, it means that securinine is a promising agent with therapeutic effect on HCC patients through inhibiting GABA receptor. B. The mechanism of thioguanine aggravating HCC. Thioguanine is a guanine analogues and it can decrease the expression of CDKN3. But, CDKN3 gene inhibits tumor growth by controlling mitosis. Hence, thioguanine may get aggravating effect on HCC patients.

Mercaptopurine(6-MP, Rank=10 in Table 2) is a medication used for cancer and autoimmune diseases [25]. As a purine analog, mercaptopurine belongs to purine antagonist anti-metabolic drugs [26]. 6-MP ribonucleotide inhibits purine nucleotide synthesis and metabolism by inhibiting an enzyme called Phosphoribosyl pyrophosphate (PRPP) amidotransferase. PRPP Amidotransferase is the rate limiting enzyme of purine synthesis [27]. This alters the synthesis and function of RNA and DNA.

Mercaptopurine interferes with nucleotide interconversion and glycoprotein synthesis. This makes the mercaptopurine can effectively inhibit the synthesis of DNA, thereby inhibiting the growth of tumor cells [28]. At present, although there is no direct experiment that mercaptopurine can inhibit the growth of HCC cells, it has been used to treat acute lymphocytic leukemia (ALL), chronic myeloid leukemia (CML), Crohn's disease, and ulcerative colitis [29]. In summary, mercaptopurine is likely to achieve a certain effect on HCC.

Reserpine(Rank=17 in Table 2) is an indole alkaloid, antipsychotic, and antihypertensive drug[30] that has been used for the control of high blood pressure and for the relief of psychotic symptoms [31]. The result of Gwak et al [32] shows that reserpine can reduce the expression level of CCND1 gene and its encoded protein. The CCND1 gene encodes the cyclin D1 protein. Cyclin D1 protein is a member of the cyclin protein family that is involved in regulating cell cycle progression. This protein plays a key role during the transition from the G1 phase, in which the cell grows, to the S phase, during which DNA is replicated. Overexpression of this protein allows cells to be easily crossed G1/S checkpoint that limits the growth of cells, which promotes tumor hyperplasia and is therefore considered to be an oncoprotein [33]. Some studies have found that CCND1 gene is over-expressed in HCC [34]. Thus, reserpine can potentially be used as an agent against HCC.

The other 11 drugs with negative $TS$ values are shown in Table 3. They are possible to aggravate HCC. 9 of them have been found having relationships with HCC in CTD database and we can infer these relationships are possibly negative. The remaining 2 drugs (Tioguanine, Rifabutin) are newly potential drugs for aggravating HCC marked as bold in Table 3. We will investigate the two drugs (Tioguanine, Rifabutin) based on PubMed.

Table 3. Eleven Aggravating Drugs for HCC in the Top-30 Drugs.

| Rank | Drug Name | Evidence | Inferred Count |
| --- | --- | --- | --- |
| 1 | Cytochalasin B | Inferred | 5 |
| 2 | Exemestane | Inferred | 2 |
| 3 | Spiperone | Inferred | 2 |
| 4 | Cinchonine | Inferred | 1 |
| 5 | Mepacrine | Inferred | 8 |
| 6 | **Tioguanine** | **NULL** | **NULL** |
| 7 | **Rifabutin** | **NULL** | **NULL** |
| 8 | N-Phenylanthranilic Acid | Inferred | 1 |
| 9 | Valinomycin | Inferred | 1 |
| 10 | Betulin | Inferred | 2 |
| 11 | Puromycin | Inferred | 13 |

Evidence represents a drug-disease association is curated, inferred or not existed in CTD database. Curated associations include three types: marker/mechanism (Evidence = "M"), therapeutic (Evidence = "T"), marker/mechanism & therapeutic (Evidence = "M&T"). If an association is inferred by CTD, Evidence = "inferred", and if it is not existed in CTD, Evidence = "NULL"; Inferred Count represents the number of inferrederence (s) for the curated and inferred associations. If an association is not existed in CTD, Inferred Count = "NULL".

Thioguanine (Rank= 6 in Table 3) is a guanine analogues, with cell cycle specificity, for the S cycle of the strongest cell sensitivity. In addition, thioguanine can inhibit the synthesis of guanosine nucleoside, by inhibiting the biological activity of guanylate kinase, the drug can inhibit the guanosine monophosphate (GMP) phosphoric acid to guanosine bisphosphate (GDP) transformation process [35]. Tioguanine is a medication used to treat acute myeloid leukemia (AML) [36], acute lymphocytic leukemia (ALL) [37], and chronic myeloid leukemia (CML) [38]. In 2005, Ganter et al. Showed that CDKN3 expression was significantly decreased after a period of administration of thioguanine [39]. The CDKN3 gene inhibits tumor growth by controlling mitosis, which is a tumor suppressor gene [40]. Dai et al. Found that CDKN3 expression in patients with HCC was significantly lower than that in normal humans. CDKN3 knockout experiments indicated that CDKN3 could inhibit tumor growth [41]. Therefore, in order to ensure the effectiveness of the treatment, clinical patients should avoid HCC patients taking thioguanine.

Rifabutin(Rank= 7 in Table 3) is a piperazine-containing rifamycin derivative, the drug has a broad spectrum of antibacterial activity. It can able to bind to the β-subunit of RNA polymerase and inhibit

RNA polymerase activity, thereby reducing the number of RNA synthesis of bacterial [42]. Rifabutin has been approved to prevent and treat disseminated infections of mycobacterium mycobacterium complex (MAC) carried by HIV-infected persons [43], and it is also used to treat multidrug-resistant tuberculosis [44]. Kobayashi et al. find that rifabutine will lead to an increase in the expression of cytochrome P450 3A4 (CYP3A4) in liver tissue [45]. CYP3A4 is an important metabolic enzyme, belongs to the cytochrome P450 family. It is also the most important component of adult liver microsomes CYP450, this gene is expressed in the intestinal, liver and kidney [46]. However, Fanni et al. find a significant increase of expression of CYP3A4 in HCC patients and overexpression of CYP3A4 gene could result in drug degradation or even a decreased therapeutic effect [47]. Therefore, for both suffering from HCC and tuberculosis patients, doctors should avoid using rifabutin to treat tuberculosis.

**Analyze potentially HCC-related Drugs through CMap database**

The CMap database [3] can not only be applied to calculate drug-disease correlations, but also can be used to identify connections between two drugs. In particular, for a same disease, if two drugs have strongly positive relationship, they may have similar effects on this disease. On the contrary, if their relationship is negative, they may have opposite effects. In this section, we further analyze the five predicted drugs (three therapeutic drugs shown in Table 1: Securinine, Mercaptopurine and Reserpine; two aggravating drugs shown in Table 2: Tioguanine and Rifabutin) based on CMap and estimate their correlations (evaluated by formula (6)) with known HCC drugs marked as "therapeutic" in CTD database [19]. The results are shown in Table 4.

**Table 4. The relationships of six predicted drugs with known HCC therapeutic drugs in CTD.**

| Predicted Drugs | Known HCC Drugs in CTD | Connectivity Scores |
|---|---|---|
| Securinine | Daunorubicin | 0.916 |

| | Troglitazone | 0.902 |
| --- | --- | --- |
| | Paclitaxel | 0.844 |
| **Mercaptopurine** | Estradiol | 0.941 |
| | Dexamethasone | 0.926 |
| | Sirolimus | 0.845 |
| | Troglitazone | 0.833 |
| **Reserpine** | Roxithromycin | 0.922 |
| | Resveratrol | 0.834 |
| Tioguanine | Genistein | -0.973 |
| | Sirolimus | -0.928 |
| | Indometacin | -0.891 |
| | Paclitaxel | -0.872 |
| Rifabutin | Calcium Folinate | -0.878 |
| | Estradiol | -0.873 |

The potentially therapeutic drugs of HCC are marked as bold. The other three drugs are potentially aggravating drugs of HCC.

For the three potentially therapeutic drugs (securinine, mercaptopurine and reserpine) marked as bold in Table 4, we find that securinine yields highly positive connectivity score (calculated by formula (6)) for daunorubicin, troglitazone and paclitaxel. Mercaptopurine is found having strongly positive relationships with estradiol, dexamethasone, sirolimus and troglitazone. Reserpine gets high positive connectivity scores with roxithromycin and resveratrol. For the two potentially aggravating drugs (tioguanine and rifabutin) in Table 4, they all have negative relationship with known HCC drugs. Tioguanine has high negative connectivity scores with genistein, sirolimus, indometacin and paclitaxel. Rifabutin have clear negative connection scores with calcium folinate and estradiol.

**Pathway Enrichment Analysis**

In this part, we analyze the relationship between these five drugs and HCC from the point of view of drug targets. First, we get the target set of drugs from DrugBank [48] because DrugBank contains the most complete information on drug and drug targets. Then, we enter these genes into DAVID [49] to obtain all the KEGG [50] pathways of the drug target. If the drug's target appears in a pathway, the pathway is picked.

From Table 5, it can be seen that the drug target of securinine and tioguanine is not clear and there is no corresponding drug target. Mercaptopurine has two drug targets, and drug targets of this drug appear in five KEGG pathways. Reserpine has two drug targets, which are included in seven KEGG pathways. Rifabutine consists of five drug targets, and nine KEGG pathways contain drug targets of the drug.

**Table 5. Pathway enrichment analysis result of six selected drugs.**

| Drug Name | Drug Targets | KEGG Pathways |
|---|---|---|
| **Securinine** | None | None |
| **Mercaptopurine** | HPRT1, PPAT | Purine metabolism; Metabolic pathways; Drug metabolism other enzymes; Alanine aspartate and glutamate metabolism; Biosynthesis of antibiotics |
| **Reserpine** | SLC18A2, SLC18A1 | Cocaine addiction; Synaptic vesicle cycle; Amphetamine addiction; Serotonergic synapse; Dopaminergic synapse; Parkinson's disease; Alcoholism |
| **Tioguanine** | None | None |
| **Rifabutin** | rpoA, rpoB, rpoC, HSP90A1, HSP90B1 | NOD-like receptor signaling pathway; **Prostate cancer;** Estrogen signaling pathway; Protein processing in endoplasmic reticulum; PI3K-Akt signaling pathway; **Pathways in cancer;** Antigen processing and presentation; Thyroid hormone synthesis; Progesterone-mediated oocyte maturation |

The potentially therapeutic drugs of HCC are marked as bold. The other three drugs are potentially aggravating drugs of HCC. "NULL" represents the drug has no targets in DrugBank at present. Thus, its corresponding KEGG pathway is "NULL" too.

In order to obtain the tissue-specific KEGG pathways of HCC, firstly, the nine genes (see Table 5) related to HCC are extended through obtaining the direct neighbors of nine genes in liver-specific protein-protein interaction (PPI) network got from GIANT [51]. Then, we obtain a subnetwork from the liver PPI network, which contains 58 genes and 838 edges with weight ≥ 0.1 (S2 Table). Finally, by using DAVID tool [49], we perform pathway enrichment analysis on these genes. The parameters of DAVID are fixed as: p-value = 0.001 and count = 5. Finally, we obtain 12 KEGG pathways [50] related to the 58 genes (see Table 6).

**Table 6. Twelve enriched KEGG pathways with HCC.**

| Pathways | Number of HCC-specific genes | P-values |
|---|---|---|
| Pathways in cancer | 20 | 3.06E-13 |
| Prostate cancer | 10 | 1.41E-08 |
| Adherens junction | 8 | 1.44E-06 |
| Endometrial cancer | 7 | 2.15E-06 |
| Colorectal cancer | 8 | 2.62E-06 |
| Apoptosis | 8 | 3.32E-06 |
| Melanoma | 7 | 1.36E-05 |
| Wnt signaling pathway | 9 | 1.41E-05 |
| Cell cycle | 7 | 3.28E-04 |
| Notch signaling pathway | 5 | 4.16E-04 |
| Basal cell carcinoma | 5 | 7.61E-04 |
| Melanogenesis | 6 | 8.61E-04 |

There are four of five pathways in which mercaptopurine drug targets exist and these four pathways have common genes with the 12 tissue-specific KEGG pathway of HCC. The interaction between the four pathways and the 12 tissue-specific KEGG pathway of HCC is shown in Fig.4. Metabolic pathways have common genes with seven pathways which are tissue-specific KEGG pathway of HCC. There are common genes between purine metabolism and one HCC related pathway. For Drug metabolism other enzymes, it exits and common genes with one HCC related pathway. There are common genes between Biosynthesis of antibiotics and one HCC related pathway. These overlap genes between the pathways of mercaptopurine drug target and HCC tissue-specific KEGG pathways have shown that mercaptopurine has a potential effect of treating HCC.

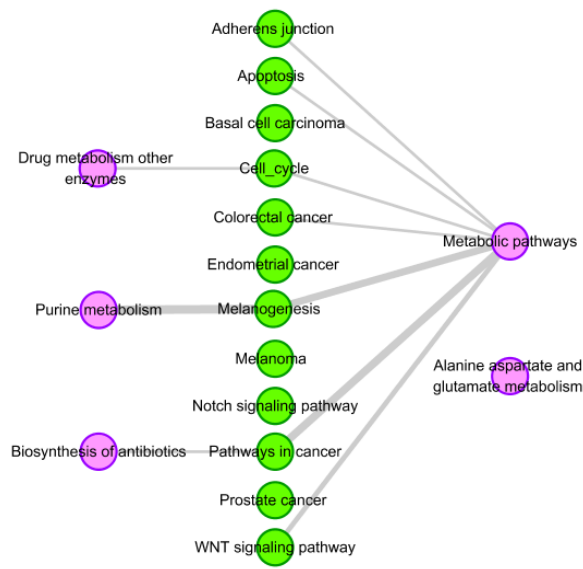

Figure 4. .Pathway analysis of mercaptopurine drug targets. The purple circle represents pathways of the mercaptopurine drug targets appeared, and the green circle represents tissue-specific KEGG pathway of HCC. The gray side indicates that there is a common gene between two pathways, and the more number of public genes is, the more rough edge is.

There are six of seven pathways in which reserpine drug targets exist and these six pathways have common genes with the 12 tissue-specific KEGG pathway of HCC. The interaction between the six pathways and the 12 tissue-specific KEGG pathways of HCC is shown in Fig.5. Cocaine addiction has common genes with five pathways which are tissue-specific KEGG pathways of HCC. There are common genes between Amphetamine addiction and seven HCC related pathways. For Serotonergic synapse, it exits and common genes with ten HCC related pathways. There are common genes between Alcoholism and seven HCC related pathways. Dopaminergic synapse has common genes with nine pathways which are tissue-specific KEGG pathways of HCC. There are common genes between Parkinson's disease and five HCC related pathways. These overlap genes between the pathways of reserpine drug target and HCC tissue-specific KEGG pathways have shown that reserpine is likely to become the treatment of liver cancer. HCC.

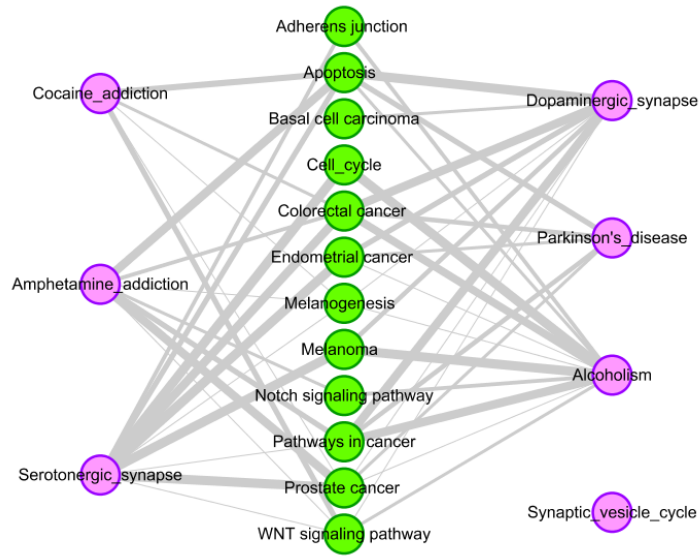

Figure 5. Pathway analysis of reserpine drug targets. The purple circle represents pathways of the reserpine drug targets appeared, and the green circle represents tissue-specific KEGG pathway of HCC. The gray side indicates that there is a common gene between two pathways, and the more number of public genes is, the more rough edge is.

Two of nine pathways in which rifabutine drug targets exist are also tissue-specific KEGG pathways of HCC highlighted in the table 4. The remaining seven pathways are associated with HCC and there are common genes with those HCC related pathways. The interaction between the six pathways and the 12 tissue-specific KEGG pathways of HCC is shown in Fig.6. There are common genes between NOD-like receptor signaling pathway and eight HCC pathways. There are common genes between Estrogen signaling pathway and nine HCC pathways. PI3K-Akt signaling pathway has common genes with eight HCC pathways. There are common genes between five different liver cancer pathways. Antigen processing and presentation pathway has common genes with three pathways of liver cancer. Thyroid hormone synthesis has common genes with five pathways of liver cancer. Progesterone-mediated oocyte maturation has common genes with ten pathways of liver cancer. There is a large number of common genes between pathways of rifabutine drug target and characteristic pathways of HCC, and it also confirms our prediction between rifabutin and HCC.

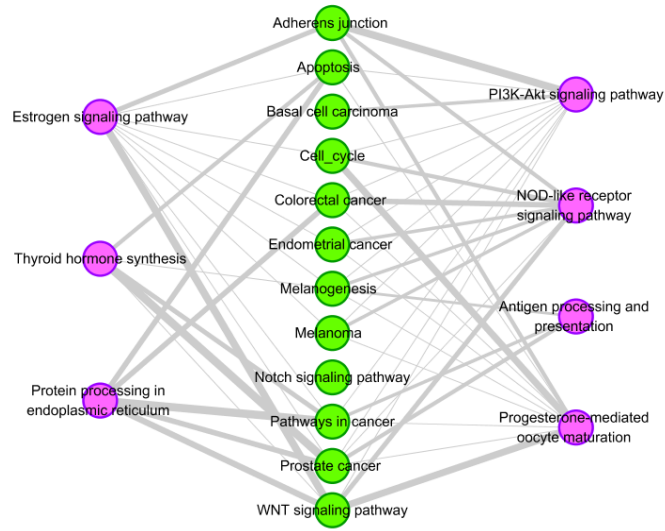

Figure 6. Pathway analysis of rifabutine drug targets. The purple circle represents pathways of the rifabutine drug targets appeared, and the green circle represents tissue-specific KEGG pathway of HCC. The gray side indicates that there is a common gene between two pathways, and the more number of public genes is, the more rough edge is.

In Table 7, we have obtained twelve KEGG pathways that are closely related to HCC. Prostate cancer pathway and Pathways in cancer are pathways in which rifabutin drug targets are appeard, and they are two of twelve KEGG pathways closely related to HCC. This suggests that our approach not only accurately predict the potential association of rifabutine, but also clearly predicts the risk of rifabutin on HCC. The other two drugs don't have the same pathway as the HCC pathway mentioned above, but there are common genes between pathways where drug targets are located and the pathways closely related to HCC.

## Methods and Materials

## Datasets

**HCC gene expression data.** The Cancer Genome Atlas (TCGA) [52] is a comprehensive and coordinated effort to accelerate our understanding of the molecular basis of cancer through the application of genome analysis technologies, including large-scale genome sequencing. TCGA researchers aims to catalogue and discover major cancer-causing genomic alterations to create a comprehensive "atlas" of cancer genomic profiles. So far, the project has analyzed large cohorts of over 30 human tumors through large-scale genome sequencing and integrated multi-dimensional analyses.

We download the gene expression profiles of HCC from TCGA, and there are 423 samples in the data set. The type of a sample is distinguished by the barcode provided by TCGA. If the fourth part of the barcode of one sample is in the range from 01 to 09, the sample is a cancer sample. If the fourth part of the barcode in the range from 10 to 19, the sample is a normal sample. The specific introduction to the barcode can be found in TCGA help file. First, we obtain gene expression matrix data (20,501×423), which contains 373 cancer samples, 50 normal samples, and 20,501 genes. Then, we standardize the expression values of all genes as follows:

$$z_{ij} = \frac{g_{ij} - mean(g_i)}{std(g_i)} \quad (1)$$

where $g_{ij}$ represents the expression value of gene $i$ in sample $j$, and $mean(g_i)$ and $std(g_i)$ respectively represent mean and standard deviation of the expression vector for gene $i$ across all samples. Finally, we use Limma [53] to analyze cancer and normal samples and get the $\log FC$ value of each gene. The definition of $\log FC$ is as follows:

$$\log FC_i = \log_2 \left( \frac{\frac{1}{|T|} \sum_{k \in T} z_{ik}}{\frac{1}{|N|} \sum_{k \in N} z_{ik}} \right) \quad (2)$$

where $\log FC_i$ is the $\log FC$ value of gene $i$; $z_{ik}$ is the normalized expression of gene $i$ in sample $k$; $T$ is the set of cancer samples ($|T|=373$), $N$ is the set of normal samples ($|N|=50$). For a gene, if its $|\log FC| \geq 1$ and $p->value \leq 0.02$, this is differentially expressed gene, or it is generally expressed gene ($|\log FC| < 1$ or $p->value > 0.02$). The thresholds of $\log FC$ and $p->value$ refer to Mark et al. [54].

**Gene expression data related to drugs.** The gene expression data related to drugs is downloaded from the CMap (http://www.broadinstitute.org/cmap/) database [3]. It contains 6,100 instances, which cover 1,309 drugs. These instances are measured on five types of human cancer cell lines, including the breast cancer epithelial cell lines MCF7 and ssMCF7, the prostate cancer epithelial cell line PC3, the nonepithelial lines HL60 (leukemia) and SKMEL5 (melanoma).

**SNP mutation data of HCC.** We download the single nucleotide polymorphism (SNP) gene mutation data of HCC from TCGA [52] database. The SNP mutation data contains 373 cancer patient sample files, and each sample file contains the detailed descriptions of all the mutated genes. Since the mutation frequency of each gene across all samples is different, we select genes with relatively high mutation frequency for further analysis. Here, the frequency is set to be no less than 11 ($11 = 373 \times 3\%$), that is a gene mutated in at least three percent of all samples. Finally, we find 406 frequently mutated genes.

## Methods

## Mutated Genes

### Defining the feature gene set of HCC

According to the data analysis we have done in section **Datasets**, we can divide the 20,501 genes into three classes based on mutation frequency and expression value. One class is kernel gene, which is mutated frequently. The second class is secondary gene, which is not mutated frequently but differentially expressed. The third class is marginal gene, which is neither mutated frequently nor differentially expressed.

In our work, we take the 406 kernel genes, i.e. frequently mutated gene, as the feature gene set of HCC.

## Calculate the therapeutic scores of drugs

We select kernel genes as the feature genes of HCC and rank them in descending order based on their expressions. For a gene, if its $\log FC$ value is greater than 0, it is stored in up-regulated gene set. Otherwise, it is stored in down-regulated gene set. Then, these kernel genes are stored in up-regulated gene set ($G_{up}$) or down-regulated gene set ($G_{down}$). Finally, we get two ordered gene lists for HCC: the up-regulated gene list and the down-regulated gene list.

We get 6,100 gene expression instances covered 1,309 drugs from CMap database. In other words, a drug may correspond to multiple instances. We rank the genes in each instance by their differential expression values between drug-treated and drug-untreated cell lines. In this way, we get 6,100 drug-related gene lists. Therefore, based on kernel genes and 6100 drug-related gene expression instances, we use a nonparametric, rank-based pattern matching strategy originally introduced by Lamb et al. [3] to evaluate the relationship between drugs and HCC (shown in Fig 1 (5)).

We take each ranked drug expression profile as reference signature and assess their similarity to HCC. we compute a connectivity score separately for the set of up- or down-regulated genes: $ES_{up}$ or $ES_{down}$. Where $n$ represents the total number of genes in the inferrederence drug expression profile; $m$ represents the size of $G_{up}$ or the size of $G_{down}$; $p$ represents the position of the input set ($p=1...m$); $V(p)$ is the position of the $p$th input gene in the gene list of drug expression profile. The computational formulas as follows [3]:

$$a = \underset{p=1}{\overset{m}{Max}} \left[ \frac{p}{m} - \frac{V(p)}{n} \right] \qquad (3)$$

$$b = \underset{p=1}{\overset{m}{Max}} \left[ \frac{V(p)}{n} - \frac{p-1}{m} \right] \qquad (4)$$

$$ES_{up/down} = \begin{cases} a_{up/down} & \text{(if } a_{up/down} > b_{up/down}) \\ -b_{up/down} & \text{(if } a_{up/down} < b_{up/down}) \end{cases} \quad (5)$$

The therapeutic score ($TS$) of a drug is calculated as follows:

$$TS = \frac{1}{k}\sum_{j=1}^{k} ES_{up} - ES_{down} \quad (6)$$

If the up-regulated genes are near the top (up-regulated) of the rank-ordered drug gene lists and the down-regulated genes are near the bottom (down-regulated) of the rank-ordered drug gene lists, we can get high positive therapeutic scores ($TS$), which indicate the drugs and HCC have similar expression profiles and the drugs might aggravate HCC. On the other hand, if the up-regulated pathway genes are near the bottom of the rank-ordered drug gene lists and the down-regulated pathway genes are near the top of the rank-ordered drug gene lists, we can get negative therapeutic scores ($TS$), which imply the given drugs and HCC have adverse expression profiles and the drugs could be treatment candidates for HCC.

## Discussions and Conclusions

We propose a method based on the combination of mutation data and differential expression data. First, we screen the collection of hepatocellular feature genes that frequently mutated in most samples of hcc based on human somatic mutation data. Then, the gene expression data of HCC in TCGA are combined to classify the gene set of HCC. Finally, the therapeutic score ($TS$) of each drug is calculated based on the kolmogorov-smirnov statistical method. By this method, five drugs associated with HCC are obtained, including three drugs that could cure HCC and two drugs that might have bad effect on HCC. There are advantages in our method. First, we take into account the essential impact of genetic changes on HCC. Secondly, we integrate multiple data to determine which type a gene should belong to. Finally, our method can clearly distinguish positive and negative relationship between drug and

HCC.

However, we discover that our approach has a strong dependence with CMap database. CMap is a database of genome-wide transcriptional data that reveals the functional relationship between drugs, genes, and diseases by manipulating small molecules of biological activity (drugs) on cultured human cells. The data contained 1309 drugs, and 6100 drug expression gene expression profile. These expression profiles were used to measure five types of human cancer cell lines, including breast cancer epithelial cell lines MCF7 and ssMCF7, prostate cancer epithelial cell lines PC3, and non-epithelial cell lines HL60 (leukemia) and SKMEL5 (melanoma). CMap database has the following shorttage. First, the number of drugs in CMap is limited, it only contains 1309 drugs, our method can't predict the drug that CMap does not contain. Secondly, the number of gene expression profiles of each drug in CMap is limited, and there will exsit measurement errors when calculate gene change. Finally, the cell line used in CMapis limited, only in the five types of human cancer cell lines are tested. Therefore, if the data in the CMap database is improved, it provides more drug expression data, our results will be more accurate and find more potential drugs for the treatment of HCC.